\documentclass[%
 reprint,
 superscriptaddress,
 amsmath,amssymb,
 aps,
]{revtex4-2}
\usepackage{amsmath, amssymb}
\usepackage{bbm}
\usepackage{graphicx}
\usepackage{dcolumn}
\usepackage{bm}
\usepackage{appendix}
\usepackage{url}
\usepackage{algorithmic}
\usepackage{algorithm}
\usepackage{amsthm}
\theoremstyle{plain}
\newtheorem{theorem}{Theorem}

\begin{document}

\title{Robust Lindbladian Tomography for Cyclic Quantum Gates}

\author{Takanori Sugiyama}
\email{sugiyama-taka@fujitsu.com}
\thanks{The author moved to affiliations 1 and 2 from affiliation 3 during the research.}
\affiliation{Quantum Laboratory, Fujitsu Limited. Nakahara-ku, Kawasaki, Kanagawa 211-8588, Japan}
\affiliation{RIKEN RQC-FUJITSU Collaboration Center, RIKEN. Wako, Saitama 351-0198, Japan.}
\affiliation{Research Center for Advanced Science and Technology, The University of Tokyo, Meguro-ku, Tokyo 153-8904, Japan.}

\date{\today}

\begin{abstract}
Precise characterization of noisy quantum operations plays an important role for realizing further accurate operations.
Quantum tomography is a popular class of characterization methods, and several advanced methods in the class use error amplification circuit (EAC), a repetition of a sequence of quantum gates, for increasing their estimation precision.
Here, we develop new theoretical tools for analyzing effects of an EAC on Lindbladian error of cyclic gates in the EAC for arbitrary finite-dimensional system, which takes non-commutativity between different gates or between ideal and error parts of a gate, periodic properties of ideal gates, and repetition of gate sequence into consideration within a linear approximation.
We also propose a tomographic protocol for the Lindbladian errors of cyclic gates based on the linear approximation, named Robust Lindbladian Tomography (RLT).
The numerical optimization at data-processing of the proposed method reduces from nonlinear to linear (positive semi-definite) programming.
Therefore, compared to the original optimization problem, the reduced one is solvable more efficiently and stably, although its numerical cost grows exponentially with respect to the number of qubits, which is the same as other tomographic methods.
\end{abstract}

\maketitle

\section{Introduction}
\begin{figure}[b]
    \centering
    \includegraphics[width=0.45\textwidth]{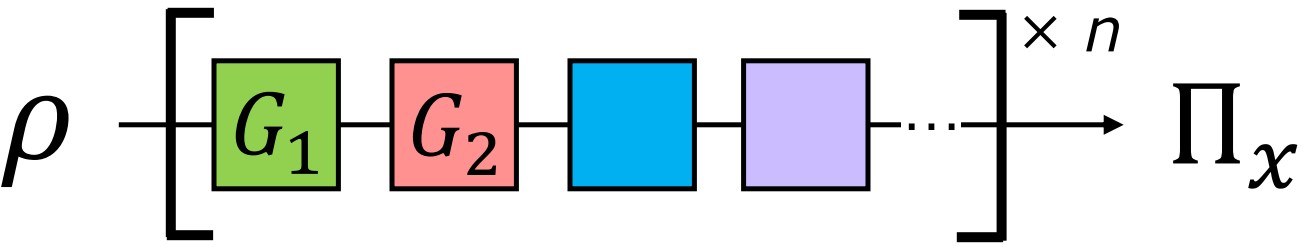}
    \caption{Quantum circuit diagram of an error amplification circuit. The superscript, ``$\times n$", means $n$ times repetition of the braketed gate sequence.}
    \label{fig:error_amplification}
\end{figure}

Further improvement of elemental quantum operations' accuracy is an inevitable task for realizing practical quantum computer.
Characterization methods such as quantum tomography \cite{Paris2004, GST_original} and randomized benchmarking \cite{Emerson2005, Emerson2007, Knill2008, Magesan2011, Magesan2012PRA, Magesan2012PRL, Gembetta2012, Chasseur2015, Wallman2016, Wallman2015, Sheldon2016, Wallman2015PRL, Cross2016, Kimmel2014} are used for improving the accuracies, and take a role to obtain information of errors of the operations.
Tomographic methods are suitable for obtaining detailed information of the errors, but its standard protocols \cite{Fano1957,Smithey1993,Hradil1997,Banaszek1999,Poyatos1997,Chuang1997,Luis1999} suffer from not-negligible systematic errors originated from mismatch of our model values on states and measurements, which is called state-preparation-and-measurement (SPAM) error.
Error amplification circuit (EAC) consists of a repetition of a sequence of quantum gates (Fig.~\ref{fig:error_amplification}).
It is used to suppress effects of such SPAM error on estimation result in advanced tomographic methods such as gate-set tomography (GST) \cite{GST}, idle tomography (IT) \cite{IT}, Hamiltonian-error amplifying tomography (HEAT) \cite{HEAT}, Floquet calibration tomography (FCT) \cite{FloquetCalibrationTomography}, and matrix-element amplification using dynamical decoupling (MEADD) \cite{MEADD}.

Such tomographic methods with EAC can be categorized into two classes by whether a method is independent of gate's implementation or not.
GST is the former, and there are no limitation on applicable types of gates. 
In addition, both of Hamiltonian error and dissipation are characterization object (states and measurements are as well).
In compensation for the high generality, it suffers high costs of experiments and data-processing.
As HEAT is for cross-resonance gates, and FCT and MEADD are for excitation-number-conserving two-qubit gates, they are the latter, and dissipation is not their characterization object. 
In return for limiting applicable types of gate and estimation object, the latter methods succeed in reducing costs of experiments and data-processing.
In the latter, however, there remain questions of whether the chosen Hamiltonian model or ignorance of dissipation is valid, although mitigation technique is used for extracting effects of dissipation on estimation of Hamiltonian error in FCT and MEADD.

In this paper, we consider a tomographic characterization of quantum gates with error amplification.
In order to reduce experimental cost, We focus on gate characterization, and we restrict the class of gates to cyclic.
Our characterization object is Lindbladian errors of cyclic gates, which include both of Hamiltonian error and dissipation.
In order to reduce numerical cost at data-processing, we introduce a linear approximation of action of EAC on the Lindbladian errors, which reduces the numerical optimization at the data-processing from nonlinear to linear (positive semi-definite) programming.
The rest of the paper is organized as follows.
Notation and settings are explained in Sec.~\ref{sec:NotationAndSettings}.
In Sec.~\ref{sec:SummaryOfTheoreticalResults}, we summarize our theoretical results; one is proposal of a tomographic protocol, and the other is a set of formulae for the linear approximation. 
Basic idea behind the protocol is explained in Sec.~\ref{subsec:BasicIdeaOfProtocol}.
Brief overview of the derived formulae is given in Sec.~\ref{subsec:OutlineOfLinearApproximations}.
Sec.~\ref{sec:conclusion} concludes the paper.
Basic knowledge on matrix representation of quantum gate and Lindbladian are given in Sec.~\ref{Sec:QuantumGateAndLindbladian}.
Detailed explanation of difficulties on theoretical analysis of EAC is given in Sec.~\ref{sec:difficulties}.
Rigorous statements of the formulae are given with their proofs in Sec.~\ref{sec:TheoreticalResults}.
A limitation of our method is discussed in Sec.~\ref{sec:LimitationOfRLT}.

\section{Notation and Settings}\label{sec:NotationAndSettings}

Let us consider arbitrary finite $d$ dimensional system.
Let $\rho \in \mathbbm{C}^{d\times d}$ and $\bm{\Pi}=\{ \Pi_x \}_{x}$ denote density matrix and POVM, respectively.
Let $\mathcal{G}$ denote a linear trace-preserving and completely-positive map representing action of a quantum gate.
Let $\bm{B}$ denote an orthonormal matrix basis on $\mathbbm{C}^{d\times d}$.
Let $|\rho \rangle\rangle$ and $G$ denote the matrix vectorization of $\rho$ and matrix representation of $\mathcal{G}$ w.r.t. $\bm{B}$.
When the gate $\mathcal{G}$ is implemented by dynamics of Lindblad master equation, $G$ is represented in the following form,
\begin{eqnarray}
    G=e^{L^{\mathrm{ideal}} + \delta L},
\end{eqnarray}
where $L^{\mathrm{ideal}}$ is the matrix representation of the ideal Lindbladian (accumulated over finite time) of the gate, and $\delta L$ is an Lindbladian error.

For a given ideal gate $\mathcal{G}^{\mathrm{ideal}}$, if there exists a positive integer k satisfying 
\begin{eqnarray}
    \left[ \mathcal{G}^{\mathrm{ideal}} \right]^k = \mathcal{I},
\end{eqnarray}
we call the smallest $k$ the period of the gate.
For example, typical quantum gates like X90 and ZX90 have period $k=4$.
We assume that the all ideal gates in the EAC has periods.
Then any gate sequence consisting the gates has a period.

In the EAC experiment depicted by Fig. 1, let $\rho$, $\bm{\Pi}$, $G$ denote actual (implemented) quantum operations that are possibly noisy.
When the superscript, $\mathrm{ideal}$ is put on an object, it corresponds to the ideal counterpart, and we assume the ideal value is known.
The difference between the actual and ideal values is its error.
In our method, the Lindbladian errors of some gates specified by an user is the estimation object, and errors of the other objects (gates, states, measurements) are not.

The probability that we observe an outcome $x$ at the experiment depicted by Fig. 1 is given as
\begin{eqnarray}
p_x (\bm{\delta L}, n) 
= \langle\langle \Pi_x | \left[ \cdots G_2 G_1 \right]^n |\rho \rangle\rangle,
\end{eqnarray}
where $\bm{\delta L}$ denote a set of Lindbladian errors for gates in the EAC.
The functionality of $p_x$ w.r.t. $\bm{\delta L}$ is highly non-linear for large $n$.
This non-linearity makes the numerical optimization for tomographic data-fitting of the model to data hard and unstable.
In order to avoid the non-linearity, we introduce linear approximations that reduce the numerical cost and instability at the data-processing.

\section{Summary of Theoretical Results}\label{sec:SummaryOfTheoreticalResults}

\subsection{Basic Idea of Protocol}\label{subsec:BasicIdeaOfProtocol}
We explain the basic idea of our tomographic protocol.
We introduce two linear approximations.
One is for gate repetition, and the other is for gate composition.

About gate repetition: Suppose that an EAC with period $k$ is given and we repeat the EAC unit $n$ times that satisfies 
\begin{eqnarray}
    n = k \cdot m + r, \label{eq:condition_on_n}
\end{eqnarray}
where $m$ is the quotient and $r$ is the residual.
Let $G^{\mathrm{unit}}$ denote the matrix representation of the repetition unit of the EAC, $L^{\mathrm{unit}, \mathrm{ideal}}$ denote the ideal Lindbladian, $dL^{\mathrm{unit}}$ denote the Lindbladian error. 
When we repeat the unit $n$ times, the unit Lindbladian error is modified as 
\begin{eqnarray}
[G^{\mathrm{unit}}]^n 
&=& e^{n L^{\mathrm{unit}, \mathrm{ideal}} + n dL^{\mathrm{unit}}} \\
&=& e^{r L^{\mathrm{unit}, \mathrm{ideal}} + f^{\mathrm{mod}}(n dL^{\mathrm{unit}})}, \label{eq:G^n_after_modulo}
\end{eqnarray}
where $f^{\mathrm{mod}}$ denote the function corresponding to the action of modulo on the ideal Lindbladian on the Lindbladian error.
The function $f^{\mathrm{mod}}$ is nonlinear, and we introduce a linear approximation of $f^{\mathrm{mod}}$ w.r.t. $dL^{\mathrm{unit}}$, which is the first linear approximation.

About gate composition: Let $G_j$ denote a matrix representation of the $j$-th gate in the EAC, $L^{\mathrm{ideal}}_j$ denote the ideal Lindbladian, and $dL_i$ denote its Lindbladian error. 
Then these matrices are related as follows.
\begin{eqnarray}
   G^{\mathrm{unit}} 
   &=& e^{L^{\mathrm{unit}, \mathrm{ideal}} + dL^{\mathrm{unit}}} \\
   &=&\prod_j G_j = \prod_j e^{L^{\mathrm{ideal}}_j + dL_j},
\end{eqnarray}
where $G_j$ can be non-commutative, and product order must obey the gate order in the EAC.
The unit Lindbladian error $dL^{\mathrm{unit}}$ depends on ideal Lindbladians $\{ L^{\mathrm{ideal}}_j \}_j$, Lindbladian errors $\{ dL_j \}_j$, and the multiplication order of the gates.
The dependency is non-linear, and we introduce a linear approximation w.r.t. $\{ dL_j \}_j$, which is the second linear approximation. 

With the two linear approximations, Eq.~(\ref{eq:G^n_after_modulo}) is rewritten as
\begin{eqnarray}
   [G^{\mathrm{unit}}]^n 
   \approx e^{r L^{\mathrm{unit}, \mathrm{ideal}} + r\cdot \sum_j f^{\mathrm{not-amp}}_j (dL_j) + n\cdot f^{\mathrm{amp}}_j (dL_j)},
\end{eqnarray}
where $f^{\mathrm{not-amp}}_j$ and $f^{\mathrm{amp}}_j$ are linear functions corresponding to not-amplified and amplified parts of $dL_j$, respectively.
With matrix logarithm, we have 
\begin{eqnarray}
   &&\ln [G^{\mathrm{unit}}]^n - r L^{\mathrm{unit}, \mathrm{ideal}} \notag \\
   &&\approx r\cdot \sum_j f^{\mathrm{not-amp}}_j (dL_j) + n\cdot \sum_j f^{\mathrm{amp}}_j (dL_j).
\end{eqnarray}
We can know $[G^{\mathrm{unit}}]^n$ by standard quantum process tomography (QPT) within some precision.
We can calculate $r$ and  $L^{\mathrm{unit}, \mathrm{ideal}}$ from $\{ L^{\mathrm{ideal}}_j \}_j$ and gate order in the EAC unit which are known information.

For simplicity of explanation, we considered single EAC above, but we can use multiple EACs for gate characterization in general.
Suppose that multiple EACs are given.
Let $a$ denote the index label for EACs and $f^{\mathrm{not-amp}}_{a, j}$ and $f^{\mathrm{amp}}_{a, j}$ denote the linear functions for $a$-th EAC on $dL_j$.
We consider the following tomographic protocol. 
\begin{enumerate}
\item[Step 1.] Construct $f^{\mathrm{not-amp}}_{a,j}$ and $f^{\mathrm{amp}}_{a,j}$ for all $a$ and $j$ (see Secs.~\ref{subsec:OutlineOfLinearApproximations} and \ref{sec:TheoreticalResults}). 
\item[Step 2.] Perform a standard QPT experiment on each EAC with the repetition number $n$.
\item[Step 3.] Calculate an estimate of a matrix representation of the action of the EAC with a data-processing method (estimator) used in standard QPT, and extract the information of its Lindbladian by matrix logarithm.
\item[Step 4.] Perform data-fitting to the extracted Lindbladian estimates with the linear model $\{ f^{\mathrm{not-amp}}_{a, j}, f^{\mathrm{amp}}_{a, j} \}_{a, j}$ with physicality constraints on optimization variables $\{ dL_j \}_j$.
\end{enumerate}
We call this tomographic protocol robust Lindbladian tomography (RLT).

By linearity of $f^{\mathrm{not-amp}}_{a, j}$ and $f^{\mathrm{amp}}_{a, j}$, the numerical optimization at Step. 4 is a positive semi-definite programming (SDP) as the standard tomographic protocols \cite{Kosut2008}.
Therefore, the original non-linear programming (NLP) is reduced to a linear programming by two linear approximations.
The key point here is that $f^{\mathrm{not-amp}}_{a, j}$ and $f^{\mathrm{amp}}_{a, j}$ are linear w.r.t. $\{ dL_j \}_j$ but they are nonlinear w.r.t. $\{ L^{\mathrm{ideal}}_j \}_j$ (and matrix logarithm is also non-linear function).
By separating the nonlinear effects of repetition and composition w.r.t. $\{ L^{\mathrm{ideal}}_j \}_j$ and calculate their (linearized) action before numerical optimization, the data-processing is reduced to SDP.

\subsection{Outline of Linear Approximations}\label{subsec:OutlineOfLinearApproximations}

As explained in Sec.~\ref{subsec:BasicIdeaOfProtocol}, out protocol uses linearized actions of gate composition and repetition.
There are four circumstances of EAC to be taken into account: (i) co-existence of Hamiltonian and dissipation dynamics, (ii) non-commutativity between different gates or between ideal and error parts of a gate, and (iii) non-linearity by multiple use of a same gate in a repetition unit, (iv) non-linearity by repetition of a gate sequence from the exponentiation form and from periodic properties of ideal gate sequence.
Detailed explanation of the circumstances are given in Sec.~\ref{sec:difficulties}.

With careful consideration on circumstances (i) to (iv), we derive mathematical formulae for arbitrary finite $d$ to analyze such linearized action of gate composition and repetition by combining an integral formula of matrix exponential derivative \cite{Wilcox1967, HayashiText}, matrix perturbation theory \cite{StewartSun1990}, and matrix diagonalization (spectral decomposition) \cite{Higham2008text}.
See Theorem \ref{theorem:composition_and_decomposition_of_matrix_exponential} in Sec.~\ref{subsec:PerturbativeFormulaeForMatrixDecomposition}, Theorem \ref{theorem:gate_composition} in Sec.~\ref{subsec:PerturbativeFormulaForCompositionOfNoisyQuantumGates}, and Theorem \ref{theorem:gate_repetition} in Sec.~\ref{subsec:PerturbativeFormulaForGateRepetition}. 
With the mathematical tools derived, we also give an algorithm for calculating the action of $ f^{\mathrm{not-amp}}_j$ and $f^{\mathrm{amp}}_j$ for a given EAC with arbitrary finite depth of circuit, which is given in Sec.~\ref{subsec:AlgorithmForCalculatingLinearMaps}.
The algorithm is applicable to a class of gates, which is wider than IT and HEAT, but is not applicable to some gates with singularity. 
Expansion of the applicability to much wider gate classes is open problem.

\section{Conclusion}\label{sec:conclusion}

We considered a tomographic characterization of Lindbladian errors of gates with error amplification circuits (EAC).
We derived perturbative formulae for obtaining linearized action of an EAC on the Lindbladian errors, which makes it possible to analyze dominant action of the EAC.
We also proposed a new protocol for the characterization with the linear approximation.
From the linearity, the numerical optimization at the data-processing is reduced from nonlinear to linear (positive semi-definite) programming, which means that the reduced optimization problem is solvable more efficiently and more stably than its original one.
We hope that the derived formulae contribute to deeper understandings of EACs and that the proposed protocol contribute to easier implementation and execution of tomographic characterization with error amplification.

\section*{Acknowlegments}
The author thanks Yoshiyasu Doi, Norinao Kouma, Ryo Murakami, and Shintaro Sato for their continued comments on the research.
This work was partially supported by JST PRESTO (JPMJPR1915), JST ERATO (JPMJER1601), and MEXT Q-LEAP (JPMXS0118068682), Japan. 

\appendix

\section{Quantum Gate and Lindbladian}\label{Sec:QuantumGateAndLindbladian}

We briefly explain basics of matrix representation of quantum gates and its generators, which are known.
See \cite{Sugiyama2021} for more detailed explanations.

Action of a quantum gate on $d$-dimensional system is represented by a linear trace-preserving (TP) and completely positive (CP) map, say $\mathcal{G}: \mathbbm{C}^{d\times d} \to \mathbbm{C}^{d\times d}$.
Let $\bm{B}$ denote an orthonormal matrix basis on $\mathbbm{C}^{d \times d} $ and $G \in \mathbbm{C}^{d^2 \times d^2}$ denote the matrix representation of $\mathcal{G}$ w.r.t. $\bm{B}$, i.e., $G:=\mathrm{HS}(\mathcal{G})$.
Its elements are defined as
\begin{eqnarray}
    G_{\alpha\beta} := \mathrm{Tr} \left[ B_{\alpha}^{\dagger} \mathcal{G} (B_{\beta}) \right], \ \alpha, \beta = 1, \ldots, d^2.
\end{eqnarray}
This is called the Hilbert-Schmidt (HS) or Pauli-Louville matrix representation.
When the system is multi-qubit system and we choose the normalized Pauli basis as $\bm{B}$, $G$ is also called the Pauli-transfer matrix (PTM).
An advantage of the HS matrix is that an action of a gate sequence is described by matrix multiplication.
For example, when we perform $\mathcal{G}_1$ first and $\mathcal{G}_2$ second, the HS matrix of the gate sequence is given as 
\begin{eqnarray}
    \mathrm{HS}(\mathcal{G}_2 \circ \mathcal{G}_1) 
    = \mathrm{HS}(\mathcal{G}_2) \mathrm{HS}(\mathcal{G}_1) 
    = G_2 G_1,
\end{eqnarray}
where $\circ$ is the composition of maps.
We use the HS matrix as the matrix representation of a quantum gate for analyzing EACs because of this property. 
Note that the order of the gate composition (and that of the matrix multiplication) is opposite to that in the quantum circuit diagram at Fig.~1.

The Hermiticity-preserving (HP) condition, which implied by the CP condition, and TP condition are equality constraints on $G$. 
When $\bm{B}$ is Hermitian matrix basis, the HP condition leads to $G\in \mathbbm{R}^{d^2 \times d^2}$.
Let $|A\rangle \rangle \in \mathbbm{C}^{d^2}$ denote the vectorization of a matrix $A \in \mathbbm{C}^{d\times d}$ w.r.t. $\bm{B}$.
The TP condition is given as
\begin{eqnarray}
   \langle \langle I_d | G = \langle \langle I_d |,
\end{eqnarray}
where $I_d$ is the $d \times d$ identity matrix.
Then, the number of its degrees of freedom reduces to $d^4 - d^2$.
The CP condition itself is an inequality condition, and it does not contribute to the reduction.
With the use of another matrix representation of $\mathcal{G}$ called Choi-Jamilkowski (CJ) matrix, the CP condition is represented simply as 
\begin{eqnarray}
    \mathrm{CJ} (\mathcal{G}) \succeq 0,
\end{eqnarray}
which is a positive semi-definite condition.

Another advantage of the HS matrix is that there is an explicit relation between a gate and its Lindbladian (accumulated over finite time).
When the quantum gate is realized by dynamics obeying the time-dependent Lindblad master equation (with $\hbar=1$),
\begin{eqnarray}
    \frac{d\rho}{dt} 
    &=& -i [H(t), \rho] \notag \\
    && + \sum_{i} \left(A_i (t) \rho A_i(t)^\dagger -\frac{1}{2} \left\{ A_i(t)^\dagger A_i(t) , \rho \right\} \right) \ \ \ \ \ \\
    &=:& \mathcal{L}_{t}(\rho), 
\end{eqnarray}
the HS matrix representation is given as 
\begin{eqnarray}
    G = \mathcal{T} \exp \left[ \int_{0}^{T}\! \! dt L_t \right],
\end{eqnarray}
where $\mathcal{T}$ is the Dyson's time-order operator, $T$ is the time period of the gate, and $L_t$ is the HS representation of the $\mathcal{L}_t$ of the Lindblad master equation. 
When $\mathcal{L}_t$ is bounded for any $t \in [0, T]$ and $T$ is finite, there exist a matrix $L \in \mathbbm{C}^{d^2 \times d^2}$ satisfying
\begin{eqnarray}
  G = e^L,
\end{eqnarray}
and this means that there exists a linear map $\mathcal{L}$ satisfying 
\begin{eqnarray}
    \mathcal{G} = e^\mathcal{L}.
\end{eqnarray}
These infinitesimal and finite-time generators are related as 
\begin{eqnarray}
    \mathcal{L} &=& \ln \left\{ \mathcal{T} \exp \left[ \int_{0}^{T}\! \! dt \mathcal{L}_t \right]  \right\}, \\
    L &=& \ln \left\{ \mathcal{T} \exp \left[ \int_{0}^{T}\! \! dt L_t \right] \right\}.
\end{eqnarray}

Conventionally, the infinitesimal generator $\mathcal{L}_t$ (and its matrix representation $L_t$) is called the Lindbladian.
In this manuscript, however, we call the finite-time generator $\mathcal{L}$ (and $L$) the Lindbladian of the gate $\mathcal{G}$ or just the Lindbladian in short, for simplicity of the terminology.
We believe this terminology does not cause any confusion, because the infinitesimal generator appears only in this section.

There exist equality and inequality constraints on the Lindbladian $\mathcal{L}$ ($L$) as there are TP and CP conditions on $\mathcal{G}$.
The equality condition on $L$ is as follows:
\begin{eqnarray}
   \Leftrightarrow \langle \langle I_d | L = \bm{0}. \label{eq:TP_L_in_HS}
\end{eqnarray}
The inequality condition is positive semi-definite condition on a submatrix of the CJ matrix of $\mathcal{J}$ as 
\begin{eqnarray}
    Q \mathrm{CJ}(\mathcal{L}) Q \succeq 0. \label{eq:CP_L_in_CJ}
\end{eqnarray}
where $Q:= (I_{d^2} - |I_{d}\rangle \rangle \langle \langle I_d |/d)$ is a projection matrix \cite{Wolf2008PRL}.
The CJ and HS matrices of a linear map are related linearly \cite{Sugiyama2021}, and Eq.~(\ref{eq:CP_L_in_CJ}) can be rewritten in the form of $L$.
Therefore, the physicality conditions on $L$ are given by Eqs.~(\ref{eq:TP_L_in_HS}) and (\ref{eq:CP_L_in_CJ}), and both are linear w.r.t. $L$.
We use this property at numerical optimization for our proposed method.

\section{Difficulties of Theoretical Analysis of EACs}\label{sec:difficulties}
Error amplification circuit (EAC) is a repetition of a sequence of quantum gates, and it is recently used in advanced tomographic characterization methods like GST, IT, and HEAT.
An EAC is used to suppress a bias on a characterization result, which originates from a mismatch of our model values on state preparation and measurement (SPAM) errors used at data-processing.
In general, effects of EAC, e.g., how and which part of gate error is amplified (or not amplified) by the given EAC, were not clear except for specific simple dynamics such as the identity gate with dissipation \cite{IT} or ZX interaction with 6-error-model on Hamiltonian without dissipation \cite{HEAT}. 
Difficulty for analyzing more general settings is mainly originated from the following four obstacles: (i) co-existence of Hamiltonian and dissipation dynamics, (ii) non-commutativity between different gates or between ideal and error parts of a gate, and (iii) non-linearity by multiple use of a same gate in a repetition unit, (iv) non-linearity by repetition of a gate sequence from the exponentiation form and from periodic properties of ideal gate sequence.

About Obstacle (i):
When we assume that a gate is realized by Hamiltonian dynamics and there are no dissipation, the action of the gate is described by a unitary matrix $U \in \mathbbm{C}^{d \times d}$ with an hermitian matrix $H \in \mathbbm{C}^{d \times d}$ as 
\begin{eqnarray}
    U = e^{-iH}.
\end{eqnarray}
We call $H$ the Hamiltonian of the gate, which is not the infinitesimal generator appearing at the Schr\"{o}dinger equation but is the geenrator accumulated over finite gate time as the termonology of ``Lindbladian'' explained in Sec.~A.
It is quite importance in quantum information experiments to characterize Hamiltonian error of a gate.
However, dissipation like energy relaxation and dephasing is also a major noise source of a gate, and it is one of the main obstacle for achieving further improvement of gate's accuracy.
So, it is better to consider both of Hamiltonian error and dissipation at the analysis of EAC.
The HS matrix representation of a gate, $G$, can treat both of them. 
Then we need to be careful about two discrepancies of $G$ and $L$ from $U$ and $H$. 
First, the size of representation matrix gets larger from $\mathbbm{C}^{d\times d}$ to $\mathbbm{C}^{d^2\times d^2}$ (or $\mathbbm{R}^{d^2\times d^2}$ with Hermitian matrix basis).
Second, the categories of these matrices differ.
A matrix $A$ is called diagonalizable if it has a decomposition of the form $A=V D V^{-1}$ with an invertible matrix $V$ and diagonal matrix $D$.
In particular, if $A$ is diagonalizable and $V$ is unitary, it is called unitarily diagonalizable.
A unitary matrix $U$ and an Hermitian matrix $H$ are unitarily diagonalizable, but $G$ and $L$ are not necessarily unitarily diagonalizable if dissipation exists.
Actually, when dissipation is strong, there is a case that $G$ and $L$ become not even diagonalizable, which is called an exceptional point \cite{Moiseyev2011text, Am-Shallem2015}.
In our analysis, in order to avoid such singularity points, we assume that the dissipation is sufficiently small and $G$ and $L$ are diagonalizable (but they are not necessarily unitarily diagonalizable).

About Obstacle (ii): 
At an unit of repetition in EAC, composition of gates (here we discuss an example of two gates for simplicity) are expressed formally as
\begin{eqnarray}
    G_2 G_1 
    = e^{L^{\mathrm{ideal}}_2 + \delta L_2} e^{L^{\mathrm{ideal}}_1 + \delta L_1} 
    =: e^{L^{\mathrm{ideal}}_{2,1} + \delta L_{2,1}},
\end{eqnarray}
where 
\begin{eqnarray}
   L^{\mathrm{ideal}}_{2,1} 
   &:=& \ln \left( e^{L^{\mathrm{ideal}}_2} e^{L^{\mathrm{ideal}}_1} \right), \\
   \delta L_{2,1}
   &:=& \ln \left( e^{L^{\mathrm{ideal}}_2 + \delta L_2} e^{L^{\mathrm{ideal}}_1 + \delta L_1}  \right) - L^{\mathrm{ideal}}_{2,1}. \ \ \ 
\end{eqnarray}
From the possible non-commutativity between $L^{\mathrm{ideal}}_2$ and $L^{\mathrm{ideal}}_1$, $L^{\mathrm{ideal}}_2$ and $\delta L_1$, $L^{\mathrm{ideal}}_1$ and $\delta L_2$, and $\delta L_2$ and $\delta L_1$, in general the composition results of Lindbladians are not just the summation of the original ones, i.e.,
\begin{eqnarray}
    L^{\mathrm{ideal}}_{2,1} 
    &\neq& L^{\mathrm{ideal}}_{2}  + L^{\mathrm{ideal}}_{1}, \\
    \delta L_{2,1} &\neq& \delta L_2 + \delta L_1. 
\end{eqnarray}
The Backer-Campbell-Hausdorff (BCH) formula is a well-knwon and standard mathematical tool for analysing such non-commutativity at composition of matrix exponential functions.
However, an approximation with the BCH series expansion up to a finite order is mathematically valid only for small matrices \cite{RossmannText}. 
At a typical setting of quantum gates used in quantum information experiments, the size of $L^{\mathrm{ideal}}$ is order of one and it is not small.
In order to analyze the non-commutativity in an appropriate way, we need a different mathematical tool.
For the purpose of that, we derive formulae based on the first order Taylor expansion of matrix exponential function with a spectral decomposition.

About Obstacle (iii):
At a gate sequence of an unit of repetition of a EAC, a gate can appear multiple times, like $[ G_1 G_2 G_1]^n$. 
This effect is nonlinear w.r.t. $G_1$, which makes the analysis of an EAC more difficult.
We propose an algorithm for calculating a dominant amplification effect of an EAC with the mathematical tool mentioned at the previous section.
The algorithm works sequentially along with the order of gates in a given gate sequence and it is designed such that it can treat cases of multiple appearance of a same gate.

About Obstacle (iv):
The repetition of a gate sequence at EAC causes high non-linearity w.r.t. gate or Lindbladian, because the order of a repetition number $n$ is typically tens to hundreds.
In addition to the non-linearity caused from the exponentiation form, there is another non-linearity, which is from a periodic property of ideal gate sequence.
Let $G^{\mathrm{ideal}}=e^{L^{\mathrm{ideal}}}$ denote the HS matrix of an unit of repetition of an EAC after composition of all gates in the unit.
When $G^{\mathrm{ideal}}$ has a period, i.e., there is a positive number $k$ satisfying $[G^{\mathrm{ideal}}]^k = I_{d^2}$, the effect of repetition at EAC has a periodic behavior because
\begin{eqnarray}
    [G^{\mathrm{ideal}}]^n &=& [G^{\mathrm{ideal}}]^r \\
    \Leftrightarrow
    e^{n L^{\mathrm{ideal}}} &=& e^{r L^{\mathrm{ideael}}}
\end{eqnarray}
holds for $n=k n_q+r$, where $n_q$ is the quotient and $r$ is the remainder.
This periodic property of the ideal gate sequence should be inherited to its noisy counterpart $[G]^n$ in some form.
Unfortunately, it is not simple like $[G]^n = e^{rL^{\mathrm{ideal}} + n \delta L}$ (wrong), because $L^{\mathrm{ideal}}$ and $\delta L$ can be non-commutative.
In more detail, the following equalities and non-equality hold:
\begin{eqnarray}
    [G]^{n} 
    &=& \left[e^{L^{\mathrm{ideal}}+\delta L} \right]^n \\
    &=& e^{n(L^{\mathrm{ideal}} + \delta L)} \\
    &=& e^{nL^{\mathrm{ideal}} + n \delta L} \\
    &\neq& e^{rL^{\mathrm{ideal}} + n \delta L}. 
\end{eqnarray}
Typically speaking, ideal gates in quantum information protocols and experiments like $X$, $Y$, $Z$, $H$, $S$, $T$, $CNOT$, $SWAP$, $X90$, $ZX90$ have period, and we cannot neglect its effect on an EAC. 
Therefore, in order to analyze the effect of the repetition, we need to take care about non-commutativity between ideal and error parts of Lindbladians again as analysis of composition of gates explained at Obstacle (ii).
For the purpose of that, we derive a formula for treating such effect based on matrix perturbation theory and a spectral decomposition.

\section{Theoretical Results}\label{sec:TheoreticalResults}

We develop theoretical tools for analyzing effects of EAC for arbitrary finite-dimensional quantum system, which is based on the first order perturbation theory with respect to Lindbladian errors of gates, taking into account all of Obstacles (i) to (iv). 
By combining the tools, we can calculate a set of matrices that quantify the dominant effects of an EAC.

Although our main target of analysis is the HS matrix of Lindbladian, the developed mathematical tools explained below hold for much larger class of diagonalizable matrices.
So, the tools are applicable to Hamiltonian as well, which would be interested in other fields of Quantum Physics.
In order to clarify such generality of our results, we use a notation $A$ and $B$ for denoting matrices. 
The matrix $A$ corresponds to an ideal matrix representation of generator of a gate, e.g., ideal Hamiltonian with imaginary factor $-iH^{\mathrm{ideal}}$ or ideal Lindbladian $L^{\mathrm{ideal}}$.
The matrix $B$ does to a sufficiently small perturbation added to $A$, representing an error on the generator, e.g., $B=-i\delta H$ or $B=\delta L$.
Then $A+B$ does to an implemented generator $-iH=-i(H^{\mathrm{ideal}} + \delta H)$ or $L=L^{\mathrm{ideal}} + \delta L$.  

We consider a square complex matrix $A, B \in \mathbbm{C}^{m \times m}$, where $m=d$ for Hamiltonian and $m=d^2$ for Lindbladian.
Suppose that $A$ is diagonalizable (not necessarily unitarily diagonalizable), and $A=\sum_i a_i P_i$ is the spectral decomposition, where $a_i$ are eigenvalues ($a_i \neq a_j$ if $i\neq j$) and $P_i$ are projections satisfying 
\begin{eqnarray}
P_i P_j = \delta_{ij} P_i \label{eq:projection_multiplication}
\end{eqnarray}
and 
\begin{eqnarray}
    \sum_j P_j = I_m .\label{eq:projection_summention} 
\end{eqnarray}

\subsection{Perturbative Formulae for Matrix Decomposition and Composition}\label{subsec:PerturbativeFormulaeForMatrixDecomposition}

We define complex values $\ell_{jk}$ with eigenvalues of $A$ as 
\begin{eqnarray}
   \ell_{jk}(A) := \left\{ 
   \begin{array}{cc}
    1 & (j=k)\\
    (e^{a_j - a_k} -1)/(a_j - a_k) & (j\neq k)
   \end{array}
   \right. .
\end{eqnarray}
Given $A$, with this $\ell_{jk}(A)$ and its projections, we define four linear maps as follows:
\begin{eqnarray}
   dcl_{A}(X) &:=& \sum_{j,k} \ell_{jk}(A) P_j X P_k , \label{def:dcl}\\
   dcr_{A}(X) &:=& \sum_{j,k} \ell_{kj}(A) P_j X P_k, \label{def:dcr}\\
   cml_{A}(X) &:=& \sum_{j,k} \frac{1}{\ell_{jk}(A)} P_j X P_k , \label{def:cml}\\
   cmr_{A}(X) &:=& \sum_{j,k} \frac{1}{\ell_{kj}(A)} P_j X P_k, \label{def:cmr}
\end{eqnarray}
where $X$ is any element of $\mathbbm{C}^{m \times m}$.
As Theorem 1 shows later, these four maps appear at the first order approximation of decomposition and composition of matrix exponential functions.
The notation, $dcl$, $dcr$, $cml$, and $cmr$, stem from abbreviations of such actions. 
\begin{itemize}
\item $dcl_{A}(X)$: \underline{d}e\underline{c}omposition of $A + X$ to the \underline{l}eft of $A$
\item $dcr_{A}(X)$: \underline{d}e\underline{c}omposition of $A + X$ to the \underline{r}ight of $A$
\item $cml_{A}(X)$: \underline{c}o\underline{m}position of $A$ and $X$ from the \underline{l}eft of $A$
\item $cmr_{A}(X)$: \underline{c}o\underline{m}position of $A$ and $X$ from the \underline{r}ight of $A$
\end{itemize}

We derived the first order perturbation formulae with these maps for decomposition and composition of matrix exponentials. 
\begin{theorem}\label{theorem:composition_and_decomposition_of_matrix_exponential}
For $A$ and $B$ mentioned above, the following decomposition formulae hold.
\begin{eqnarray}
   e^{A+B} &=& e^{\mathrm{dcl}_{A}(B)}e^{A} + O(\| B \|^2), \label{eq:dcl} \\
   e^{A+B} &=& e^{A} e^{\mathrm{dcr}_{A}(B)} + O(\| B \|^2), \label{eq:dcr} 
\end{eqnarray}
Additionally, if $A$ satisfies 
\begin{eqnarray}
e^{a_j - a_k} \neq 1, \ \forall j, k \ (j\neq k), \label{eq:condition_not_singular}
\end{eqnarray}
the following composition formulae hold.
\begin{eqnarray}
   e^{B} e^{A} = e^{A + cml_{A}(B)} + O(\| B \|^2), \label{eq:cml}\\
   e^{A} e^{B} = e^{A + cmr_{A}(B)} + O(\| B \|^2), \label{eq:cmr}
\end{eqnarray}
where $\| \cdot \|$ denote the Frobenius norm.
\end{theorem}
We give the proof of Theorem 1 in Sec.~\ref{sec:proof_theorem_1}.

A major difference between our result (Theorem 1) and the BCH formula is a treatment of series expansion orders of $A$ and $B$.
In the BCH formula, its series expansion order is for both of $A$ and $B$, but int our result the order ris only for $B$ (up to the 1st order).
Our result can be interpreted as the BCH formula with order of $A$ up to the infinity and that of $B$ up to one. 
Additionally, in the BCH formula, both of $A$ and $B$ are assumed to be sufficiently small.
There are several sufficient condition for the convergence of BCH series expantion, and an explicit inequality \cite{RossmannText} is  
\begin{eqnarray}
\| A \| + \| B \| \le \ln 2 \approx 0.693147\ldots . \label{eq:BCH_Sufficient_Condition}
\end{eqnarray}
Unfortunately, typical quantum gates like $\pi/4$-, $\pi/2$-, and $\pi$-pulse gates do not satisfy Eq.~(\ref{eq:BCH_Sufficient_Condition}).
For example, in the cases of ideal Lindbladians of 1-qubit X90 gate ($A=L^{\mathrm{ideal}}_{X90}$) and 2-qubit ZX90 gate ($A=L^{\mathrm{ideal}}_{ZX90}$), which is a popular 2-qubit gate used at fixed-frequency superconducting qubit \cite{Blais2021}) are 
\begin{eqnarray}
\| A \| = \left\{
\begin{array}{cc}
   \pi / \sqrt{2} \approx 2.22144\ldots & (\mbox{X90}) \\
   \sqrt{2} \pi \approx 4.44288\ldots & (\mbox{ZX90})
\end{array}
\right. ,
\end{eqnarray}
and both exceed the R.H.S. of Eq.~(\ref{eq:BCH_Sufficient_Condition}).
On the other hand, in our result the size of $A$ is arbitrary.
We numerically evaluated the relative approximation errors of Eqs.~(\ref{eq:dcl}) to (\ref{eq:cmr}) for 1-qubit, 2-qubit, and 1-qutrit gates, and compared Eqs.~(\ref{eq:cml}) and (\ref{eq:cmr}) with the BCH formula.
These examples indicate that our result is more appropreate for treating typical quantum gates than the BCH formula.

\subsection{Proof of Theorem 1}\label{sec:proof_theorem_1}
We give the proof of Theorem 1.
Let $X(t) \in \mathbbm{C}^{m \times m}$ denote a complex square matrix that is parametrized with $t$ smoothly.
Then the following integral formulae for derivative of the matrix exponential of $X(t)$ hold \cite{Wilcox1967, HayashiText}.
\begin{eqnarray}
\frac{d e^{X(t)}}{dt}  
&=& \int_0^1 \!\!\! \!ds\ e^{s X(t)} \frac{dX(t)}{dt} e^{(1-s)X(t)} \label{eq:matrix_exponential_derivative_1}\\
&=& \int_0^1 \!\!\! \!ds\ e^{(1-s) X(t)} \frac{dX(t)}{dt} e^{sX(t)}. \label{eq:matrix_exponential_derivative_2}
\end{eqnarray}
Note that $X(t)$ is not necessarily diagonalizable at Eqs.~(\ref{eq:matrix_exponential_derivative_1}) and (\ref{eq:matrix_exponential_derivative_2}), although we combine diagonalizability requirement with them in the proof below.

   First, we derive Eqs.~(\ref{eq:dcl}) and (\ref{eq:dcr}).
   We consider the case of $X(t) = A + t B/\|B\|$.
   The Taylor expansion of $e^{X(t)}$ at $t=0$ up to the first order is given as  
   \begin{eqnarray}
      e^{A + tB/\| B\|} = e^{A} 
      + \left. \frac{d e^{X(t)}}{dt}  \right|_{t=0} \cdot t 
      + O(|t|^2). \label{eq:taylor_expanstion_of_X(t)}
   \end{eqnarray}
   With Eq.~(\ref{eq:matrix_exponential_derivative_1}), $X(0)=A$, and $d X(t) / dt |_{t=0}=B/\|B\|$, the first order term in the RHS of Eq.~(\ref{eq:taylor_expanstion_of_X(t)}) is rewritten as 
   \begin{eqnarray}
    &&\left. \frac{d e^{X(t)}}{dt} \right|_{t=0} \cdot t \notag \\
     &=& \left.  \int_0^1 \!\!\! \!ds\ e^{sX(t)} \frac{dX(t)}{dt} e^{-sX(t)} \cdot e^{X(t)} \right|_{t=0} \cdot t \\
     &=& \left.  \int_0^1 \!\!\! \!ds\ e^{sX(0)} \frac{dX(t)}{dt} \right|_{t=0} e^{-sX(0)} \cdot e^{X(0)}  \cdot t \ \ \\
     &=& \int_0^1 \!\!\! \!ds\ e^{s A} B e^{-s A} \cdot e^{A} \frac{t}{\|B\|}. \label{eq:derivative_X(t)_line3}
   \end{eqnarray}
   The line integral in Eq.~(\ref{eq:derivative_X(t)_line3}) is solvable with the spectral decomposition of A as 
   \begin{eqnarray}
      &&\int_0^1 \!\!\! \!ds\ e^{s A} B e^{-s A} \notag \\
      &=& \int_0^1 \!\!\! \!ds\ \left( \sum_{j} e^{s a_j} P_j \right) B \left( \sum_{k} e^{-s a_k} P_k \right) \\
      &=& \sum_{j, k} \int_0^1 \!\!\! \!ds\ e^{s (a_j - a_k)} P_j B P_k \\
      &=& \sum_{j, k} \ell_{j k}(A) P_j B P_k = dcl_A (B), \label{eq:after_integral}
   \end{eqnarray}
   where we used the following equalities,
   \begin{eqnarray}
      \int_0^1 \!\!\! \!ds\ e^{s (a_j - a_k)}
      &=& \left\{ 
         \begin{array}{cc}
            1 & \mbox{if}\ j = k \\
            \frac{e^{a_j - a_k} -1}{a_j - a_k} & \mbox{if}\ j \neq k
         \end{array}
      \right. \\
      &=& \ell_{jk}(A).
   \end{eqnarray}
   By substituting $t = \|B\|$ into Eqs.~(\ref{eq:taylor_expanstion_of_X(t)}) and (\ref{eq:derivative_X(t)_line3}), and combining them with Eq.~(\ref{eq:after_integral}), we obtain
   \begin{eqnarray}
      e^{A + B} 
      &=& e^{A} + dcl_A(B) e^{A} + O(\|B\|^2) \\
      &=& \left( I + dcl_A(B) \right) e^{A} + O(\| B \|^2) \\
      &=& e^{dcl_A(B)} e^{A} + O(\| B \|^2),
   \end{eqnarray}
   which is Eq.~(\ref{eq:dcl}).
   Similarly, with Eq.~(\ref{eq:matrix_exponential_derivative_2}), we obtain Eq.~(\ref{eq:dcr}). $\square$
   
   Second, we derive Eqs.~(\ref{eq:cml}) and (\ref{eq:cmr}).
   Suppose that A satisfies $e^{a_j - a_k} \neq 1$ for any $j$ and $k$ if $j\neq k$.
   Then all $\ell_{jk}(A)$ are non-zero, each $1/\ell_{jk}(A)$ takes finite value, and the definition of $cml_A$ by Eq.~(\ref{def:cml}) makes sense.
   With Eq.~(\ref{eq:dcl}) and the linearity of $cml_A$, the first term of the RHS of Eq.~(\ref{eq:cml}) is rewritten as 
   \begin{eqnarray}
       e^{A + cml_A(B)} 
       &=& e^{dcl_A \circ cml_A (B)} e^A + O(\| cml_A(B) \|^2) \ \ \ \ \ \ \\
       &=& e^{dcl_A \circ cml_A (B)} e^A + O(\| B \|^2). \label{eq:proof_cml_1}
   \end{eqnarray}
   From Eqs.~(\ref{eq:projection_multiplication}) and (\ref{eq:projection_summention}), for any $X \in \mathbbm{C}^{m \times m}$,
   \begin{eqnarray}
       && dcl_A \circ cml_A (X)\notag \\
       &=& \sum_{j,k} \ell_{jk}(A) P_j \left( \sum_{j^{\prime}, k^{\prime}} \frac{1}{\ell_{j^\prime k^\prime}(A)} P_{j^\prime} X P_{k^\prime} \right) P_k \ \ \ \ \ \\
       &=& \sum_{j,k} \sum_{j^{\prime}, k^{\prime}} \frac{\ell_{jk}(A)}{\ell_{j^\prime k^\prime}(A)} P_j P_{j^\prime} X P_{k^\prime} P_k \\
       &=& \sum_{j,k} \sum_{j^{\prime}, k^{\prime}} \frac{\ell_{jk}(A)}{\ell_{j^\prime k^\prime}(A)} \delta_{j j^{\prime}} P_j X \delta_{k k^{\prime}}P_k \\
       &=& \sum_{j,k} P_j X P_k \\
       &=& \left( \sum_j P_j \right) X \left( \sum_k P_k \right) \\
       &=& X
   \end{eqnarray}
   hold.
   Therefore we have
   \begin{eqnarray}
       dcl_A \circ cml_A = id, \label{eq:dcl_cml_id}
   \end{eqnarray}
   where $id$ is the identity map (this indicates $cml_A = (dcl_A)^{-1}$).
   With Eq.~(\ref{eq:dcl_cml_id}), Eq.~(\ref{eq:proof_cml_1}) is rewritten as
   \begin{eqnarray}
      e^{A + cml_A(B)} 
      = e^{B} e^A + O(\| B \|^2),
   \end{eqnarray}
   which is equivalent to Eq.~(\ref{eq:cml}).
   Similarly, we have
    \begin{eqnarray}
       dcr_A \circ cmr_A = id, \label{eq:dcr_cmr_id}
   \end{eqnarray}  
   and we obtain an equation equivalent to Eq.~(\ref{eq:cmr}).
   $\square$

   As far as the author knows, there are two known results related to the proof of Theorem 1.
   The first one is about Eqs.~(\ref{eq:matrix_exponential_derivative_1}) and (\ref{eq:matrix_exponential_derivative_2}).
   Let us define a map $ad$ as 
   \begin{eqnarray}
      ad_{X}(Y) := [X, Y] = XY - YX, \ \forall X, Y \in \mathbbm{C}^{m \times m}.
   \end{eqnarray}
    The derivative of matrix exponential on a smoothly parametrized matrix $X(t)$ is represented with $ad$ as 
   \begin{eqnarray}
   \frac{d e^{X(t)}}{dt}  
   &=& e^{X(t)} \frac{1 - e^{-ad_{X(t)}}}{ad_{X(t)}} \frac{d X(t)}{dt} \label{eq:matrix_exponential_derivative_3} \\
   &=& e^{X(t)} \sum_{k=0}^{\infty} \frac{(-1)^k}{(k+1)!} (ad_{X(t)})^k \left( \frac{d X(t)}{dt} \right), \label{eq:matrix_exponentail_derivative_4}\ \ \ \ \ 
   \end{eqnarray}
   which are well knownn in the field of Lie algebra \cite{RossmannText}.
   Eqs.~(\ref{eq:matrix_exponential_derivative_1}) and (\ref{eq:matrix_exponential_derivative_2}) are treating the same quantity as Eqs.~(\ref{eq:matrix_exponential_derivative_3}) and (\ref{eq:matrix_exponentail_derivative_4}), and they are equivalent in that sense.
   We used the integral formulae (Eqs.~(\ref{eq:matrix_exponential_derivative_1}) and (\ref{eq:matrix_exponential_derivative_2})) in order to derive Eqs.~(\ref{eq:dcl}) and (\ref{eq:dcr}) because we found that the integration in the integral formulae is solvable by combining with the spectral decomposition and that we can obtain a closed form with finite summation, which is an advantage over infinite series in Eq.~(\ref{eq:matrix_exponentail_derivative_4}). 

   The second one is a closed form of $d e^{X(t)}/ dt |_{t=0}$ with $X(t)=A + t B$ for Hermitian $A$ and $B$ \cite{Deledalle2022}.
   In contrast to the use of the integral formulae (Eqs.~(\ref{eq:matrix_exponential_derivative_1}) and (\ref{eq:matrix_exponential_derivative_2})) here, the derivation relies on some specific properties of Hermitian matrices, and the result of \cite{Deledalle2022} is applicable to neither cases of Lindbladians ($A+B=L$) nor cases of Hamiltonians ($A+B = -i H$, which is anti-Hermitian).
   On the other hands, the proof of Theorem 1 is valid for diagonalizable $A$ and general (small) $B$, which contains cases of diagonalizable Lindbladians, cases of Hamiltonian (anti-Hermitian), Hermitian, and unitary.

\subsection{Perturbative Formula for Composition of Noisy Quantum Gates}\label{subsec:PerturbativeFormulaForCompositionOfNoisyQuantumGates}

By combining Eqs.~(\ref{eq:dcl}), (\ref{eq:dcr}), (\ref{eq:cml}), and (\ref{eq:cmr}), which are for ideal and error parts of generator of a gate, we obtain a composition formula for two noisy gates.
We still continue to use the notation $A, B \in \mathrm{C}^{m \times m}$, where $A=L^{\mathrm{ideal}}$ and $B=\delta L$ with $m=d^2$ for Lindbladian, and $A=-i H^{\mathrm{ideal}}$ and $B=-i \delta H$ with $m=d$ for Hamiltonian.
\begin{theorem}\label{theorem:gate_composition}
Suppose that $A$ and $A^{\prime} \in \mathbbm{C}^{m \times m}$ are diagonalizable and satisfy Eq.~(\ref{eq:condition_not_singular}).
Let $B$ and $B^{\prime}$ are sufficiently small perturbations added to $A$ and $A^{\prime}$, respectively.
Let $C$ denote the ideal generator of the composed gate, i.e., 
\begin{eqnarray}
C:=\ln ( e^A e^{A^{\prime}} ).
\end{eqnarray}
Then, the following composition formula holds.
\begin{eqnarray}
   e^{A + B} e^{A^{\prime} + B^{\prime}}
   = e^{C + cml_{C} \circ dcl_{A}(B) + cmr_{C} \circ dcr_{A^{\prime}} (B^{\prime})} \notag \\
   + O(\|B\|^2, \|B\| \|B^{\prime}\|, \|B^{\prime}\|^2). \label{eq:composition_two_gates}
\end{eqnarray}
\end{theorem}
Theorem 2 clarifies how generator errors $B$ and $B^{\prime}$ are transformed by the composition of two gates up to the first order of them, i.e.,
\begin{eqnarray}
    B &\to& cml_{C} \circ dcl_{A}(B), \\
    B^{\prime} &\to& cmr_{C} \circ dcr_{A^{\prime}} (B^{\prime}).
\end{eqnarray}
We can treat composition of more than two gates by applying Eq.~(\ref{eq:composition_two_gates}) to a gate sequence recursively. 

\hspace{2pt}\\
{\bf Proof of Theorem 2}:\\
Each terms in the LHS of Eq.~(\ref{eq:composition_two_gates}) is decomposed as
\begin{eqnarray}
    e^{A+B} &=& e^{dcl_A(B)}e^{A} + O(\|B\|^2), \\
    e^{A^{\prime} + B^{\prime}} &=& e^{A^{\prime}} e^{dcr_{A^{\prime}}(B^{\prime})} + O(\| B^{\prime} \|^2).
\end{eqnarray}
With these equations, the LHS of Eq.~(\ref{eq:composition_two_gates}) is rewritten as
\begin{eqnarray}
    && e^{A+B} e^{A^{\prime} + B^{\prime}} \notag \\
    &=& e^{dcl_A(B)}e^{A} e^{A^{\prime}} e^{dcr_{A^{\prime}}(B^{\prime})} + O(\|B\|^2 , \| B^{\prime} \|^2) \\
    &=& e^{dcl_A(B)}e^{C} e^{dcr_{A^{\prime}}(B^{\prime})} + O(\|B\|^2 , \| B^{\prime} \|^2) \\
    &=& e^{C+cml_{C}\circ dcl_A(B)} e^{dcr_{A^{\prime}}(B^{\prime})} + O(\|B\|^2 , \| B^{\prime} \|^2) \ \ \ \ \ \ \ \\
    &=& e^{C + cml_{C} \circ dcl_{A}(B) + cmr_{C} \circ dcr_{A^{\prime}} (B^{\prime})} \notag \\
   && + O(\|B\|^2, \|B\| \|B^{\prime}\|, \|B^{\prime}\|^2),
\end{eqnarray}
where, at the last equality, we used 
\begin{eqnarray}
    cmr_{C+\delta C} (X) = cmr_{C}(X) + O(\|\delta C \| \| X \|) \label{eq:cmr_delta_on_base}
\end{eqnarray}
with $\delta C=cml_{C}\circ dcl_A(B)$ and $X=dcr_{A^{\prime}}(B^{\prime})$.
Eq.~(\ref{eq:cmr_delta_on_base}) is derived with the fact that effects of sufficiently small perturbation on eigenvalues and projections are the first order, i.e., $a_j(C+\delta C) = a_j(C) + O(\| \delta C \|)$, $P_j (C+\delta C)=P_j (C) + O(\| \delta C \|)$.
$\square$

\subsection{Perturbative Formula for Gate Repetition}\label{subsec:PerturbativeFormulaForGateRepetition}

We introduce tools for analyzing effect of gate repetition on generator error.
In order to clarify the effect of gate repetition taking into account the non-commutativity explained at Obstacle (iv) in Sec.~\ref{sec:difficulties}, we introduce two linear maps.
\begin{eqnarray}
   ssp_{A}(B) &:=& \sum_j P_j B P_j, \label{eq:ssp}\\
   sspc_{A}(B) &:=& \sum_{j,k (j\neq k)} P_j B P_k. \label{eq:sspc}
\end{eqnarray}
From Eq.~(\ref{eq:projection_summention}), 
\begin{eqnarray}
    ssp_A + sspc_A = id_m \label{eq:ssp_sspc_id}
\end{eqnarray}
holds.

The effect of gate repetition is described by $ssp$ and $sspc$ as the follows. 
\begin{theorem}\label{theorem:gate_repetition}
Let $A, B \in \mathbbm{C}^{m \times m}$.
Suppose that $A$ and $A+B$ are diagonalizable respectively, and $\|B\|$ is sufficiently small.
We assume that $e^A$ has period, denoted by $k$. 
Let $n=k n_q + r$ denote a repetition number of a gate, which is a positive integer.
The integer $n_q$ is the quotient, and $r$ is the remainder. 
Then the following equality holds,
\begin{eqnarray}
   \left[ e^{A+B} \right]^n = e^{r A + r\cdot sspc_{A}(B) + n\cdot ssp_{A}(B)}
   + O(\| B \|^2). \label{eq:repetition_theorem}
\end{eqnarray}
\end{theorem}
Theorem \ref{theorem:gate_repetition} clarifies that the amplified part of the error $B$ by the repetition, which is proportional to $n$, is $ssp_{A}(B)$, and non-amplified part is $sspc_{A}(B)$.
The former corresponds to change of eigenvalues by addition of error $B$, and the later does to that of eigenvectors.

Let us consider an interpretation of Theorem 3 in cases of quantum gates.
Roughly speaking, dominant effect of a gate is a rotation of a (generalized) Bloch vector if we neglect dissipation.
Action of rotation is characterized by rotation axis and rotation angle.
When we perform the gate repeatedly, the rotation is executed many times.
Suppose that the ideal action of the rotation has period, and there are errors on the axis and angle.
The angle error is accumulated by increasing the repetition number until the accumulated value exceeds $\pm 2\pi$ (or $\pm\pi$).
However, the axis error is not accumulated by the increase of repetition, because the rotation axis (with error) is invariant under the rotation.
The angle error corresponds to change of eigenvalues of the generator, and the axis error does to that of eigenvectors of the generator.
Eq.~(\ref{eq:repetition_theorem}) quantifies such rough picture on the action of repetition within the first order approximation, which is also applicable to cases that the Hamiltonian dynamics and dissipation co-exist.

\hspace{2pt}\\
{\bf Proof of Theorem 3}:\\
Let us define $C:=A+B$.
By assumption, $C$ and $A$ are diagonalizable.
We use the following notation.
\begin{eqnarray}
    C &=& V_C D_C V_C^{-1}, \\
    A &=& V_A D_A V_A^{-1},
\end{eqnarray}
where we choose $V_C$ satisfying $V_C = V_A +O(\|B\|))$ to eliminate degrees of freedom on invertible matrices in the diagonalization \cite{Higham2008text}.
Let us define $\delta D_C$ as
\begin{eqnarray}
    \delta D_C := D_C - D_A,
\end{eqnarray}
which is a diagonal matrix containing changes of eigenvalues by small perturbation $B$. 
Then we rewrite $C=A+B$ as follows:
\begin{eqnarray}
    C
    &=& A+B \label{eq:A+B}\\
    &=& V_{C} D_{C} V_{C}^{-1} \\
    &=& V_{C} (D_{A} + \delta D_{C} )V_{C}^{-1} \\
    &=& V_{C} D_{A} V_{C}^{-1} + V_{C} \delta D_C V_C^{-1} \\
    &=& \tilde{A} + \tilde{B}, \label{eq:A+B_tilde}
\end{eqnarray}
where we define
\begin{eqnarray}
\tilde{A} &:=& V_C D_A V_C^{-1}, \\
\tilde{B} &:=& V_C \delta D_C V_C^{-1}.
\end{eqnarray}
Note that $\tilde{A}$ and $\tilde{B}$ are commutative.
With the commutativity, we have
\begin{eqnarray}
\left[ e^C \right]^n 
&=& \left[ e^{A+B} \right]^n \\
&=& \left[ e^{\tilde{A} + \tilde{B}} \right]^n \\
&=& e^{n\tilde{A} + n\tilde{B}} \\
&=& e^{n\tilde{A}} e^{n\tilde{B}}. \label{eq:proof_repetition_1}
\end{eqnarray}
The first term in Eq.~(\ref{eq:proof_repetition_1}) is rewritten as
\begin{eqnarray}
e^{n \tilde{A}}&=& e^{(k n_q + r) \tilde{A}} \\
&=& e^{r\tilde{A}}, \label{eq:nAtilde_to_rAtilde}
\end{eqnarray}
which is because eigenvalues of $A$ and $\tilde{A}$ are same, and they have the same period $k$.

The matrix $\tilde{B}$ represents change of eigenvalues with perturbed eigenspace. 
With $V_C = V_A + O(\| B \|)$ and $\delta D_C = O(\| B \|)$, we have
\begin{eqnarray}
    \tilde{B}
    &=& V_C \delta D_C V_C^{-1} \\
    &=& V_A \delta D_C V_A^{-1} + O(\| B \|^2) \\
    &=& ssp_{A} (B) + O(\| B \|^2), \label{eq:Btilde_ssp}
\end{eqnarray}
where we used known result on change of eigenvalues by sufficiently small perturbation \cite{StewartSun1990},
\begin{eqnarray}
V_A \delta D_C V_A^{-1} = ssp_{A} (B) + O(\| B \|^2).    
\end{eqnarray}
By combining Eqs.~(\ref{eq:A+B}), (\ref{eq:A+B_tilde}), (\ref{eq:Btilde_ssp}), and (\ref{eq:ssp_sspc_id}), we have
\begin{eqnarray}
\tilde{A}
&=& A + B - \tilde{B} \\
&=& A + B - ssp_{A} (B) + O(\| B \|^2)  \\
&=& A + (id_m - ssp_{A}) (B) + O(\| B \|^2) \\
&=& A + sspc_{A} (B) + O(\|B \|^2). 
\end{eqnarray}
Therefore, with Eqs.~(\ref{eq:nAtilde_to_rAtilde}) and (\ref{eq:Btilde_ssp}), Eq.~(\ref{eq:proof_repetition_1}) is rewritten as
\begin{eqnarray}
\left[ e^C \right]^n 
&=& e^{r\tilde{A}} e^{n \tilde{B}} \\
&=& e^{r\tilde{A} + n\tilde{B}} \\
&=& e^{r A + r \cdot sspc_A(B) + n \cdot ssp_A (B)} +O(\|B\|^2). \ \ \ \ \ \ \
\end{eqnarray}
$\square$

\subsection{Algorithm for calculating linear maps}\label{subsec:AlgorithmForCalculatingLinearMaps}

An EAC consists of a sequence of gates and its repetition.
Therefore, by combining Theorems 2 and 3, we can clarify how and which part of a generator error is amplified or not-amplified.
Such effect is characterized by six linear maps defined by Eqs.~(\ref{eq:dcl}), (\ref{eq:dcr}), (\ref{eq:cml}), (\ref{eq:cmr}), (\ref{eq:ssp}), and (\ref{eq:sspc}).
The action of a linear map can be described by its matrix representation, and the matrix representation is useful for the analysis of EAC because composition of maps becomes matrix multiplication in the representation.

Before explaining the details of the algorithm, we introduce more detailed notations.
Suppose that we can use $n_g$ independent gates in an EAC experiment, and that each gate is labeled with an integer $i \in \{ 1, \ldots , n_g \}$.
Let $G_i^{\mathrm{ideal}}$ and $G_i$ denote the ideal and actual (noisy) HS matrix representations of the $i$-th gate, respectively. 
Let $L^{\mathrm{ideal}}_i$ and $\delta L_i$ denote the ideal and error Lindbladians of the $i$-th gate, i.e.,
\begin{eqnarray}
    G_i^{\mathrm{ideal}} &=& e^{L^{\mathrm{ideal}}_i} , \\
    G_i &=& e^{L^{\mathrm{ideal}}_i + \delta L_i}.
\end{eqnarray}
Let $n_u$ denote the number of gates in the repetition unit of gate sequence in the EAC.
Then the ideal and actual HS matrix for the unit, $G^{\mathrm{unit}, \mathrm{ideal}}$ and $G^{\mathrm{unit}}$, is given as 
\begin{eqnarray}
    G^{\mathrm{unit}, \mathrm{ideal}} &=& G_{i_{n_u}}^{\mathrm{ideal}} \cdots G_{i_2}^{\mathrm{ideal}} G_{i_1}^{\mathrm{ideal}}, \\
    G^{\mathrm{unit}} &=& G_{i_{n_u}} \cdots G_{i_2} G_{i_1},
\end{eqnarray}
where $i_j \in \{ 1, \ldots , n_g \}$ is the gate label for $j$-th gate in the unit ($j=1, \ldots, n_u$).

Let $L^{\mathrm{unit}, \mathrm{ideal}}$ denote the ideal Lindbladian of the repetition unit, i.e.,
\begin{eqnarray}
   L^{\mathrm{unit}, \mathrm{ideal}}
   := \ln G^{\mathrm{unit}, \mathrm{ideal}}.
\end{eqnarray}
Let $\mathcal{F}^{\mathrm{unit}}_i$ ($i=1, \ldots , n_g$) denote the linear maps for the linearized action of the gate composition of the repetition unit on each Lindbladian errors $\delta L_i$.
Then the following first order approximation of $G^{\mathrm{unit}}$ holds.
\begin{eqnarray}
   G^{\mathrm{unit}}
   &=& \exp \left[ L^{\mathrm{unit}, \mathrm{ideal}} + \sum_{i=1}^{n_g} \mathcal{F}^{\mathrm{unit}}_{i} (\delta L_i) \right] \notag \\
   &&+ O(\| \delta L_i \|^2 ).
\end{eqnarray}
We show an algorithm for calculating $\{ \mathcal{F}^{\mathrm{unit}}_i \}_{i=1}^{n_g}$ in Fig.~\ref{fig:algorithm_maps_gate_composition_in_unit}, which is based on Theorem 2. 
With Theorem 3, we have 
\begin{eqnarray}
    f^{\mathrm{not-amp}}_i 
    &=& sspc_{L^{\mathrm{unit}, \mathrm{ideal}}} \circ \mathcal{F}^{\mathrm{unit}}_{i}, \ \ \ \ \ \ \ \\
    f^{\mathrm{amp}}_i
    &=& ssp_{L^{\mathrm{unit}, \mathrm{ideal}} } \circ \mathcal{F}^{\mathrm{unit}}_{i}. \label{eq:Fi1}
\end{eqnarray}

\begin{figure}[!t]
\begin{algorithm}[H]
    \caption{Calculate maps $\{ \mathcal{F}^{\mathrm{unit}}_i \}_{i=1}^{n_g}$}
    \label{alg1}
    \begin{algorithmic}[1]
    \REQUIRE $\{ L^{\mathrm{ideal}}_i \}_{i=1}^{n_g}$, $\{ i_j \}_{j=1}^{n_u}$
    \ENSURE $\{ \mathcal{F}^{\mathrm{unit}}_i \}_{i=1}^{n_g}$
    \FOR{$i=1, \ldots , n_g$}
        \STATE $\mathcal{F}_i \leftarrow 0$
    \ENDFOR
    \STATE $A^\prime \leftarrow L^{\mathrm{ideal}}_{i_1}$
    \STATE $\mathcal{F}_{i_1} = id$
    \FOR{$j = 2, \ldots , n_u$}
        \STATE $A \leftarrow L^{\mathrm{ideal}}_{i_j}$
        \STATE $C \leftarrow \exp (e^A e^{A^\prime} )$
        \IF{$C$ is not singular}
            \FOR{$i=1, \ldots , n_g$}
                \STATE $\mathcal{F}_i \leftarrow cmr_C \circ dcr_{A^\prime} \circ \mathcal{F}_i$
            \ENDFOR    
            \STATE $\mathcal{F}_{i_j} \leftarrow \mathcal{F}_{i_j} + cml_C \circ dcl_A$
        \ELSE
            \STATE stop        
        \ENDIF
    \ENDFOR
    \RETURN $\{ \mathcal{F}_{i} \}_{i=1}^{n_g}$
    \end{algorithmic}
\end{algorithm}
\caption{Algorithm for calculating the linear maps for the linearized action of gate composition in a repetition unit.}
\label{fig:algorithm_maps_gate_composition_in_unit}
\end{figure}

In RLT, there are several linear maps like $dcl$, $dcr$, $cml$, $cmr$, $ssp$, $sspc$, $\mathcal{F}^{\mathrm{unit}}_i$, $f^{\mathrm{not-amp}}_i$, and $f^{\mathrm{amp}}_i$. 
Direct uses of such maps themselves may not be suitable for programming implementation of RLT.
The author used their matrix representations at the Python implementation.
The maps act on an HS matrix of Lindbladian, meaning that they are maps from $\mathbbm{C}^{d^2 \times d^2}$ to $\mathbbm{C}^{d^2 \times d^2}$.
Their matrix representation with an orthonormal matrix basis on $\mathbbm{C}^{d^2 \times d^2}$ is in $\mathbbm{C}^{d^4 \times d^4}$.
By reflecting the HP of Lindbladian into the basis, we can reduce the size to $\mathbbm{R}^{d^4 \times d^4}$.
As in the case of matrix representation of quantum gates, the composition of maps, which is used in Algorithm 1 (Fig.~\ref{fig:algorithm_maps_gate_composition_in_unit}) and in Eqs.~(\ref{eq:Fi1}), are replaced with matrix multiplication.

\section{Limitation of RLT}\label{sec:LimitationOfRLT}

In RLT, we assume that all gates used in an EAC have period.
So, RLT is not applicable to characterization of gates with no period, which can appear in quantum variational algorithms.
Additionally, in Algorithm 1 (Fig.~\ref{fig:algorithm_maps_gate_composition_in_unit}), $cml$ and $cmr$ are used, which means we assume Eq.~(\ref{eq:condition_not_singular}).

We found that ideal Lindbladians for rotation of 180 degrees do not satisfy Eq.~(\ref{eq:condition_not_singular}) and RLT is not applicable to them.
Such singular class of gates includes important gates in quantum computation like $X$, $Y$, $Z$, $CNOT$, $SWAP$.
One possible solution of this singularity problem is to implement such 180-degree-rotation gates with 90-degree-rotation gates, because 90-degree-rotation gates like $X90$, $Y90$, $Z90$, and $ZX90$ satisfy Eq.~(\ref{eq:condition_not_singular}), and RLT is applicable to them.
For example, $X$ and $CNOT$ can be implemented by $[X90]^2$ and $[Z_1 X_2 90] [Z_1 90]^{-1} [X_2 90]^{-1}$.
Characterization of a 180-degree-rotation gate can be done indirectly by combining characterization results of its 90-degree-rotation components with RLT.
We note that, in addition to 90-degree rotation gates, the identity gate and $T$ gate satisfy Eq.~(\ref{eq:condition_not_singular}), and RLT is applicable to them.
So, the applicable range of RLT is wider than IT (the identity gate only) and HEAT (ZX90 gate only).

In the solution, gate implementation is affected by a limitation of characterization method (RLT).
Of course, it is better not to have such limitation.
Improvement of the current version of RLT as applicable to such singular gates is an open problem.
The origin of the singularity is non-invertibility of $dcl$ and $dcr$.
This is related to a singularity of the derivative of exponential map in Lie algebra, and such singularity point is called a critical point in Dirrerential Geometry \cite{GallierQuaintanceText}.
Mathematical knowledge on critical points might be useful for solving this problem of RLT.

\end{document}